\documentstyle[preprint,floats,epsf,tighten,aps]{revtex}
\newcommand{\be}{\begin{equation}}
\newcommand{\ee}{\end{equation}}
\newcommand{\bea}{\begin{eqnarray}}
\newcommand{\eea}{\end{eqnarray}}

\def\del{\partial}

\def\dag{\dagger}

\begin{document}
\draft
\preprint{ \parbox{1.5in}{ \leftline{WM-01-103}
                            \leftline{JLAB-THY-01-07}
                          }
          }
\title{ The stability of the scalar $\chi^2\phi$ interaction}
\author{ Franz Gross$^{1,2}$, \c{C}etin \c{S}avkl{\i}$^{1}$, and John 
Tjon$^{3}$  }
\address{
$^1$Department of Physics, College of William and Mary, Williamsburg,
Virginia 23187\\
$^2$Jefferson Lab,
12000 Jefferson Avenue, Newport News, VA 23606 \\
$^3$Institute for Theoretical Physics, University of Utrecht, Princetonplein
5, P.O. Box 80.006, 3508 TA Utrecht, the Netherlands.}

\date{\today}
\maketitle
\begin{abstract}

A scalar field theory with a $\chi^\dag\chi\phi$ interaction is known
to be unstable. Yet it has been
used frequently without  any sign of instability in standard text
book examples and research articles.  In order to reconcile these
seemingly conflicting results, we show that the theory is stable if
the Fock space of all intermediate states is limited to a {\em
finite} number of $\chi{\bar\chi}$ loops associated with field $\chi$
that appears quadradically in the interaction, and that instability
arises only when intermediate states include these loops to all
orders.
 
\end{abstract}
\pacs{11.10St, 11.15.Tk}



Scalar field theories with a $\chi^\dag\chi\phi$ interaction (which we
will subsequently denote simply by $\chi^2\phi$)  have been used
frequently without any sign of instability, despite a proof in 1959 by
G.~Baym~\cite{Baym} showing that the theory is unstable.  For example, it
is easy to show that, for a limited range of coupling values $0\le g^2\le
g^2_{crit}$, the simple sum of bubble diagrams for the propagation of a
single $\chi$ particle leads to a stable ground state, and it is shown in
Ref.~\cite{SAVKLI1} that a similar result also holds for the {\it exact\/}
result in ``quenched'' approximation.  However, if the scalar
$\chi^2\phi$ interaction is unstable, then this instability should be
observed even when the coupling strength $g$ is vanishingly small
$g^2\rightarrow 0^+$, as pointed out recently by Rosenfelder and
Schreiber\cite{ROSY} (see also Ref.~\cite{Dar}).  Both the simple 
bubble summation
and the quenched  calculations do not exhibit this behavior.  Why do the simple
bubble  summation and the exact quenched calculations produce stable 
results for a
finite range of coupling values?

A clue to the answer is already provided by the simplest
semiclasical estimate of the ground state energy.  In this  approximation the
gound state energy is obtained by minimizing
\be
E_0=m^2\chi^2 + \frac{1}{2}\mu^2 \phi^2 -g\phi\chi^2\, ,
\ee
where $m$ is the bare mass of the matter particles, and $\mu$ the mass
of the ``exchanged'' quanta, which we will refer to as the {\it
mesons\/}.  The minimum occurs at
\be
E_0=m^2\chi^2 - g^2\frac{\chi^4}{2\mu^2}\, .
\ee
The ground state is therefore stable (i.e. greater than zero) provided
\be
g^2<g^2_{\rm crit}=\frac{2m^2\mu^2}{\chi^2}\, .
\ee
This simple estimate suggests that the theory is stable over a
limited range of couplings {\it if the strength of the $\chi$ field is
finite\/}.  In this letter we develop this argument more precisley and show
under what conditions it holds.

We start in the Heisenberg representation, where the
fields depend on time and the states are independent of time.  The fields
are expanded in terms of creation and annihilation operators
\bea
\chi(t,\bf{r})&=&\int d\tilde{k}_m \left[
a(k)\,e^{-ik\cdot x} + b^{\dag}(k)
\,e^{ik\cdot x}\right]  \nonumber\\
\phi(t,\bf{r})&=&\int d\tilde{k}_\mu \left[
c(k)\,e^{-ik\cdot x} + c^{\dag}(k)
\,e^{ik\cdot x}\right] \label{fields}
\eea
where $x=\{t, {\bf r}\}$ and
\be
d\tilde{k}_m \equiv \frac{ d^3k } { (2\pi)^3\,2E_m(k) }
\ee
with $E_m(k)=\sqrt{m^2+k^2}$. The equal-time commutation relations are
\be
\left[a(k),a^\dag(k')\right] = (2\pi)^3\,2E_m(k)\, \delta^3(k-k')\, .
\ee

The Lagrangian for the $\chi^2\phi$ theory is
\be
{\cal L}=\chi^\dag\left[\del^2-m^2+g\phi\right]\chi+\frac{1}{2}\,
\phi\left(\del^2-\mu^2\right)\phi\, ,
\label{lagr0}
\ee
and the hamiltonian $H$ is a normal ordered product of interacting (or dressed)
fields $\phi_d$ and $\chi_d$
\bea
H[\phi_d,\chi_d,t\,]
= \int d^3r\, : \Biggl\{
\left(\frac{\del\chi_d}{\del t}\right)^2+({\bf\nabla}\chi_d)^2+m^2 \chi_d^2
\nonumber\\
+\frac{1}{2}\biggl[ \left(\frac{ \del \phi_d }{\del t}
\right)^2+(\vec{\nabla}\phi_d)^2+\mu^2\phi_d^2 \biggr]
- g \chi_d^2 \phi_d \Biggr\}:\, . \label{hamiltonian}
\eea
This hamitonian conserves the {\it difference\/} between
number of matter and the number of antimatter
particles, which we denote by $n_0$.  Eigenstates of the hamiltonian
will therefore be denoted by
$\left|n_0,\lambda\right>$, where $\lambda$ represents the other quantum
numbers that define the state. Hence, allowing for the fact that the eigenvalue
may depend on the time,
\bea
  H[\phi_d,\chi_d,t\,] \left|n_0,\lambda\right> = M_{n_0,\lambda}(t)
\left|n_0,\lambda\right>\, . \label{eveqn}
\eea

In the absence
of an exact solution of (\ref{eveqn}), we may estimate it from the
equation
\bea
M_{n_0,\lambda}(t)=&&\left<n_0,\lambda\right| H[\phi_d,\chi_d,t\,]
\left|n_0,\lambda\right>\nonumber\\
=&& \left<n_0,\lambda\right|U^{-1}(t,0) H[\phi,\chi,0\,]U(t,0)
\left|n_0,\lambda\right>\nonumber\\
\equiv&& \left<n_0,\lambda, t\right| H[\phi,\chi,0\,]
\left|n_0,\lambda,t\right> \, ,\label{matrixelem}
\eea
where $U(t,0)$ is the time translation operator which carries the 
hamiltonian from
time $t=0$ to later time $t$.  We have also chosen $t=0$ to be the 
time at which
the interaction is turned on, $\phi_d(t)=U^{-1}(t,0)\phi(0)U(t,0)$, 
and the last
step simplifies the discussion by permitting us to work with a hamiltonian
constructed from the {\it free\/} fields $\phi$ and $\chi$.  [If the 
interaction
were turned on at some other time
$t_0$, we would obtain the same result by absorbing the additional phases
$\exp(\pm i Et_0)$ into the creation and annhilation operators.]

At $t=0$ the hamiltonian in normal order reduces to
\bea
H[\phi,&&\chi,0\,] =\int d\tilde{k}_m\,E_m(k)\,{\cal N}_0(k,k)\nonumber\\
+&&\int
d\tilde{p}_\mu\,E_\mu(p)\, c^\dag(p)c(p) \nonumber\\
-&&\frac{g}{2}\int\,\frac{d\tilde{k}_m\,d\tilde{k'}_m}
{\omega(k-k')}{\cal N}_1(k,k')\biggl[c^\dag(k'-k) +c(k-k')\biggr]
\eea
where
\bea
&&{\cal N}_0(k,k')=\left\{a^\dag(k)a(k')+b^\dag(k)b(k')\right\}\nonumber\\
&&{\cal N}_1(k,k')={\cal 
N}_0(k,k')+\left\{a^\dag(k)b^\dag(-k')+a(-k)b(k')\right\}
\eea
and
$\omega(k)=\sqrt{\mu^2+{\bf k}^2}$.   To evaluate the matrix element
(\ref{matrixelem}) we express the the eigenstates as a sum of free 
particle states
with $n_0$ matter particles, $n_{\rm pair}$ pairs of $\chi{\bar\chi}$ 
particles,
and
$\ell$ mesons:
\bea
\left|n_0,\lambda,t\right>\equiv&&
\left|n_0,\alpha(t),\beta(t)\right>\nonumber\\
=&&\frac{1}{\gamma(t)}\,\sum_{n_{\rm pair}=0}^\infty
\sum_{\ell=0}^\infty\alpha_{n_{\rm pair}}(t)
\beta_\ell(t)\left|n_0,n_{\rm pair},\ell\right> \label{general}
\eea
where $\gamma(t)$ is a normalization constant (defined below), the time
dependence of the states is contained in the time dependence of the
coefficients $\alpha(t)$ and $\beta(t)$,  and
\bea
\left| n_{0},n_{\rm pair},\ell\right>\equiv \int
\frac{\left|k_1,\cdots,k_{n_1};q_1,\cdots,q_{n_2};p_1,\cdots,p_\ell\right>}
{\sqrt{(n_0+n_{\rm pair})!\,n_{\rm pair}!\,\ell!}}
\label{fockstate}
\eea
with $n_1=n_0+n_{\rm pair}$, $n_2=n_{\rm pair}$ and
\be
\int=\int\,\prod_{i=1}^{n_1}
\,d\tilde{k_{i}}\,f(k_i)\;\prod_{j=1}^{n_2}\,
d\tilde{q_j}\,f(q_j)\;\prod_{l=1}^{\ell}\,
d\tilde{p_l}\,g(p_l)
\ee
The particle masses in $d\tilde{k}$ and $d\tilde{p}$ have been
suppressed; their values should  be clear from the context.
The normalization of the functions $f(p)$ and $g(p)$ is chosen to be
\be
\int d\tilde{k} \, f^2(k)=\int d\tilde{p}\, g^2(p)\equiv 1
\ee
which leads to the normalization
\bea
\left<n'_{0},n'_{\rm pair},\ell'\,|\,n_{0},n_{\rm pair},\ell
\right>=&&\delta_{n'_{0},n_{0}}\,\delta_{n'_{\rm pair},n_{\rm
pair}}\, \delta_{\ell',\ell} \nonumber\\
\left<n_{0},\lambda,t\,|\,n_{0},\lambda,t
\right>=&&1  \, ,
\eea
if $\gamma(t)=\alpha(t)\beta(t)$ with
\bea
\alpha^2(t)=&&\sum_{n_{\rm pair}=0}^{\infty} \alpha_{n_{\rm
pair}}^2(t) = \alpha(t)\cdot\alpha(t)  \nonumber\\
\beta^2(t)=&&\sum_{\ell=0}^{\infty}\,\beta_{\ell}^2(t) =
\beta(t)\cdot\beta(t)
\, .
\eea
The expansion coefficients
$\{\alpha_{n_{\rm pair}}(t) \}$ and $\{\beta_\ell(t)\}$ are vectors in
infinite dimensional spaces.

In principle the scalar cubic interaction in four dimensions requires
ultraviolet regularization. However the issue of regularization and
the question of stabilty are qualitatively unrelated. For example, the
cubic interaction is also unstable in dimensions lower than four, where
there is  no need for regularization. The ultraviolet regularization
would  have an effect on the behavior of functions $f(p)$, and $g(p)$,
which  are left  unspecified in this discussion except for their
normalization.

The matrix element (\ref{matrixelem}) can now be evaluated.  Assuming that
$f(k)=f(-k)$ and $g(k)=g(-k)$, it becomes:
\bea
M_{n_0,\lambda}(t)
&&=\left\{n_0+2L(t)\right\}\tilde{m} +G(t)\,\tilde{\mu}\nonumber\\
&&-g V \left\{n_0+2L(t) +2L_1(t)\right\}\,
\sqrt{G_1(t)} \, , \label{generalelement}
\eea
where the constants $\tilde{m}$, $\tilde{\mu}$, and $V$ are
\bea
\tilde{m}&\equiv&\int\,d\tilde{k}\,E_m(k)\,f^2(k),\ \ 
\tilde{\mu}\equiv\int\,d\tilde{p}\,E_\mu(p)\,g^2(p)\nonumber\\
V&\equiv&\int\,\frac{d\tilde{k}_m\,d\tilde{k}'_m\,
f(k)f(k')g(k- k')} {\sqrt{m^2+({\bf k}-{\bf k}')^2}}\, ,
\eea
and the time dependent quantities are
\bea
L(t)=&&\sum_{n_{\rm pair}=0}^\infty
\frac{n_{\rm pair}\,\alpha^2_{n_{\rm pair}}(t)}{\alpha^2(t)},\ \ 
G(t)=\sum_{\ell=0}^\infty\frac{\ell\,\beta_\ell^2(t)}{\beta^2(t)}\nonumber\\ L_1(t)=&&\sum_{n_{\rm
pair}=1}^\infty
\frac{\sqrt{n_0+n_{\rm pair}}\sqrt{n_{\rm
pair}}\;\alpha_{n_{\rm pair}}(t)\,
\alpha_{{n_{\rm pair}}-1}(t)} {\alpha^2(t)}
\nonumber\\
\sqrt{G_1(t)}=&&\sum_{\ell=1}^\infty \frac{\sqrt{\ell}\;
\beta_\ell(t)\beta_{\ell-1}(t)}{\beta^2(t)} \, . \label{averages}
\eea
Note that $L$ and $G$ are the {\it average\/} number of matter pairs 
and mesons,
respectively, in the intermediate state.

The variational principle tells us that the correct mass must be
equal to or larger than (\ref{generalelement}).  This inequality may be
simplified by using the Schwarz inequality to place an upper limit on the
quantities $L_1$ and $G_1$.  Introducing the vectors
\bea
f_1=&&\{\alpha_1,\sqrt{2}\,\alpha_2,\cdots\}=\{\sqrt{n}\;\alpha_{n}\}
\nonumber\\
f_2=&&\{\sqrt{n_0+1}\,\alpha_0,\sqrt{n_0+2}\,\alpha_1,\cdots \}
=\{\sqrt{n_0+n}\;\alpha_{n-1}\}\nonumber\\
h=&&\{\beta_1,\sqrt{2}\,\beta_2,\cdots
\} =\{\sqrt{\ell}\;\beta_{\ell}\}\, ,
\eea
we may write
\bea
&&L_1(t)=\frac{f_1(t)\cdot f_2(t)}{\alpha^2(t)}\le
\frac{\sqrt{f_1^2(t)\,f_2^2(t)}}{\alpha^2(t)}\nonumber\\
&&\hspace{3.2cm}=\sqrt{L(t)\{n_0+1+L(t)\}}\nonumber\\
&&\sqrt{G_1(t)}=\frac{h(t)\cdot \beta(t)}{\beta^2(t)}\le
\frac{\sqrt{h^2(t) \beta^2(t)}}{\beta^2(t)}=\sqrt{G(t)}\, . \label{conditions}
\eea
Hence, suppressing explicit reference to the time dependence of $L$ and $G$,
Eq.~(\ref{generalelement}) can be written
\bea
M_{n_0,\lambda}(t)
\ge&&\left(n_0+2L\right)\tilde{m} +G\,\tilde{\mu}\nonumber\\
&&-g V
\left\{\left(\sqrt{n_0+1+L}+\sqrt{L}\right)^2-1\right\}\,
\sqrt{G} \, . \label{phivariation}
\eea
Minimization of the ground state energy with respect to the average number of
mesons $G$ occurs at
\be
\sqrt{G_0}=\frac{gV}{2\tilde{\mu}}\,\left\{\left(\sqrt{n_0+1+L}+
\sqrt{L}\right)^2-1\right\} \, .
\ee
At this minimum point the ground state energy is bounded by
\bea
M_{n_0,\lambda}(t)\geq&& \left\{n_0+2L\right\}\tilde{m}-\tilde{\mu} G_0\, .
\label{variational}
\eea
This result shows that the ground state is stable for
couplings in the interval $0<g^2<g_{crit}^2$ with
\be
g_{crit}^2\equiv\frac{4\,\tilde{\mu}\,\tilde{m}\,
(n_0+2L)}{V^2\left\{\left(\sqrt{n_0+1+L}+\sqrt{L}\right)^2-
1\right\}^2} \, .
\label{gcrit}
\ee
This interval is nonzero if the number of matter particles, $n_0$, and
the average number of $\chi\bar{\chi}$ pairs, $L$, is finite.  In
particular, {\it if there are no $Z$ diagrams or $\chi\bar{\chi}$ loops
in the intermediate states, then the ground state will be stable for a
limited range of values of the coupling.\/}

This result also suggests strongly that the system is unstable when
$g^2>g^2_{\rm crit}$, or when $L\to\infty$ (implying that $g^2_{\rm
crit}\to0$).  However, since Eq.~(\ref{variational}) is only a lower
bound, our argument does not provide a proof of these latter assertions.

To strengthen our understanding of the causes of instability in a
$\chi^2\phi$ theory, we turn to the Feynman-Schwinger representation 
(FSR).  This
can be used to show that the ground state is (i) {\it stable\/} when
$Z$-diagrams are included in intermediate states, but (ii) {\it 
unstable\/} when
matter loops are included.

The FSR is a path integral approach for finding the exact result  for
propagators in field theory. It replaces integrals over fields by
integrals over all possible covariant trajectories of the
particles\cite{SIMONOV1}.  It has been applied to the $\chi^2\phi$ 
interaction in
Refs.~\cite{SAVKLI1,TJON1,TJON2,SAVKLI0,SAVKLI2,SAVKLI3}.

The covariant trajectory $z(\tau)$ of the  particle
is parametrized as a function of the proper time $\tau$. In $\chi^2\phi$
theory the FSR  expression for the 1-body propagator for a dressed
$\chi$-particle in quenched approximation in Euclidean space is given by
\begin{eqnarray}
G(x,y)&=&\int_0^{\infty} ds \left[\frac{N}{4\pi s}\right]^{2N}
\;\prod^{N-1}_{i=1}\int d^4z_i\nonumber\\
&&\times\exp\biggl\{-K[z,s]-V[z,s_r]\biggr\}\, , \label{FS}
\end{eqnarray}
where the integrations are over all possible particle trajectories
(discretized into $N$ segments with $N-1$ variables $z_i$ and boundary
conditions $z_0=x$, and $z_N=y$) and the kinetic and self energy
terms are
\bea
K[z,s]&=& m^2s+\frac{N}{4s}\sum_{i=1}^{N}(z_i-z_{i-1})^2\, ,\label{kin}\\
V[z,s]&=&-\frac{g^2s^2}{2N^2}\sum_{i,j=1}^{N}
\Delta\left(\delta z_{ij},\mu\right)\,  ,
\label{pot}
\eea
where $\Delta(z,\mu)$ is the Euclidean progagator of the meson
(suitably regularized), $\delta 
z_{ij}=\frac{1}{2}(z_i+z_{i-1}-z_j-z_{j-1})$, and
\be
s_r\equiv \frac{s}{1+(s-s_0)^2/\Gamma^2}\, ,
\ee
(The substitution $s\to s_r$ does not alter the results, but is necessary
to correctly transform the original integral from Minkowsky space to
Euclidean space, where it can be numerically evaluated.  For a detailed
discussion of this technical point, see Ref.~\cite{SAVKLI1}.)

In preparation for a discussion of the effects of $Z$-diagrams and
loops, we first discuss the stability of Eq.~(\ref{FS}) when neither 
$Z$-diagrams
nor loops are present.  To make the discussion explicit, consider the one
body propagator in 0+1 dimension.  Since the integrals converge, we make the
crude approximation that each $z_i$ integral is approximated by {\it one\/}
point (since we are excluding $Z$-diagrams, the points may lie along 
the classical
trajectory).  If the boundary conditions are
$z_0=0$ and
$z_N=T$ the points along the classical trajectory are $z_i=iT/N$, and
\begin{equation}
K[z,s]= m^2s+\frac{N}{4s}\sum_{i=1}^{N}(z_i-z_{i-1})^2=m^2s+\frac{T^2}{4s}\, .
\label{KE}
\end{equation}
If the interaction is zero, this has a stationary point at $s=s_0=T/(2m)$,
giving
\be
K[z,s]=K_0=mT\, ,
\ee
yielding the expected free particle mass $m$.  [Note
that {\it half\/} of this result comes from the sum over
$(z_i-z_{i-1})^2$.]  The potential term (\ref{pot}) may be similarily
evaluated; it gives a negative contribution that reduces the mass.

\begin{figure}
\begin{center}
\mbox{
    \epsfxsize=2.0in
\epsffile{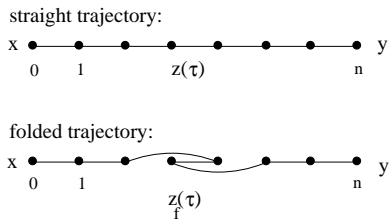}
}
\end{center}
\caption{It is possible to create particle-antiparticle pairs using
folded  trajectories. However folded trajectories are suppressed by the
kinematics.}
\label{folded.fig}
\end{figure}

We now turn to a discussion of the effect of $Z$-diagrams.  For the simple
estimate of the kinetic energy, Eq.~(\ref{KE}), we chose
integration points $z_i=iT/N$ uniformly spaced along a line.  The classical
trajectory connects these points without doubling back, so that they increase
monotonically with proper time, $\tau$.  However, since the 
integration over each
$z_i$ is independent, there also exists trajectories where $z_i$ does 
not increase
monotonically with $\tau$.  In fact, for every choice of integration 
points $z_i$
there exist trajectories with $z_i$ monotonic in $\tau$ and trajectories with
$z_i$ non-monotonic in $\tau$.  The latter double back in time, and describe
$Z$-diagrams in the path integral formalism.   Two such trajectories that pass
through the {\it same\/} points $z_i$ are shown in 
Fig.~\ref{folded.fig}.  These
two trajectories contain the same points $z_i$, but ordered in 
different ways, and
both occur in the path integral.

Now, since the total self energy is the sum of potential
contributions $V[z,s]$ from all $(z_i,z_j)$  pairs, irrespective of how these
coordinates are ordered, it must be the same for the straight 
trajectory $z(\tau)$
and the folded trajectory $z_f(\tau)$:
\be
V[z_f,s]=V[z,s]\, .
\ee
However, according to Eq.~(\ref{kin}), the  kinetic energy of the folded
trajectory is larger than the kinetic energy of  the straight trajectory
\be
K[z_f,s]>K[z,s]\, ,
\ee
because it includes some terms with larger values of $(z_i-z_{i-1})^2$.  Since
the kinetic energy term is always positive, the folded trajectory
($Z$-graph) is always suppressed (has a larger exponent) compared with a
corresponding  unfolded trajectory (provided, of course, that $g^2<g^2_{\rm
crit}$).

\begin{figure}
\begin{center}
\mbox{
    \epsfxsize=1.5in
\epsffile{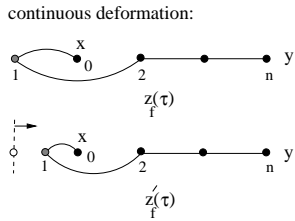}
}
\end{center}
\caption{A folded trajectory at the end point of the path, and a similar one
with $z_1$ closer to $z_0$.}
\label{folded2.fig}
\end{figure}

This argument holds only for cases where the trajectory does {\it not\/} double
back to times {\it before\/} $z_0=0$ or {\it after\/} $z_N=T$.  An example of
such a trajectory is shown in Fig.~\ref{folded2.fig} (upper panel).  Here we
compare this folded trajectory to another folded trajectory, $z'_f$, with point
$z_1$ {\it closer\/} to the starting point $z_0$ (lower panel of
Fig.~\ref{folded2.fig}).  This new folded trajectory has points spaced closer
together, so that the kinetic energy is smaller and the potential energy is
larger, and therefore
\be
K[z_f,s]-V[z_f,s] > K[z'_f,s]-V[z'_f,s]\, .
\ee
It is clear that the larger the folding in the trajectory, the less 
energetically
favorable is the path, and the most favorable path is again an 
unfolded trajectory
with no points outside of the limits $z_0<z_i<z_N$.

While these arguments have been stated in 0+1 dimensions for 
simplicity, they are
not dependent on the number of dimensions, and hold for the realistic 
case of 3+1 dimensions.

We conclude that a calculation in quenched
approximation, where the  creation of particle-antiparticle pairs can only come
from $Z$-graphs, must be {\it more\/} stable (produce a larger mass) 
than a similar
calculation without {\it any\/}
Z-graphs or $\chi\bar{\chi}$ pairs.  The quenched
$\chi^2\phi$ theory therefore is bounded by the same limits given in
Eq.~(\ref{gcrit}).  This conclusion supports, and is supported by, 
the results of
Refs.~\cite{SAVKLI1,SAVKLI2,SAVKLI3} which show, in the quenched
approximation, that the $\chi^2\phi$ interaction  is stable for a 
finite  range of
coupling strengths.

It is now clear that the instability of $\chi^2\phi$ theory must be
due to either (i) the possibility of creating an infininte number of closed
$\chi\bar{\chi}$ {\it loops\/}, or (ii) the presence of an infinite number of
matter particles (as in an infinite medium).  Indeed, the original proof
given by Baym used the possibility of loop creation from the vacuum 
to prove that
the vacuum was unstable. In fact, the FS representation can be used to show
explicitly that the critical coupling
$g^2_{\rm crit}$ decreases as
$1/L$, where $L$ is the number of closed loops, in agreement with the estimate
of Eq.~(\ref{gcrit})\cite{FS3}.

This work was supported in part by the US Department of Energy under grant
No.~DE-FG02-97ER41032. The Southeastern Universities Research 
Association (SURA)
operates the Thomas Jefferson National Accelerator Facility under DOE contract
DE-AC05-84ER40150.

\end{document}